# The emergence of global phase coherence from local pairing in underdoped cuprates

Shusen Ye[1], Changwei Zou[1], Hongtao Yan[2], Yu Ji[1], Miao Xu[1], Zehao Dong[1], Yiwen Chen[2], Xingjiang Zhou[2] and Yayu Wang[1,3,4]*

[1]State Key Laboratory of Low Dimensional Quantum Physics, Department of Physics, Tsinghua University; Beijing 100084, P. R. China

[2]Beijing National Laboratory for Condensed Matter Physics, Institute of Physics, Chinese Academy of Sciences; Beijing 100190, P. R. China

[3]New Cornerstone Science Laboratory, Frontier Science Center for Quantum Information , Beijing 100084, P. R. China

[4]Hefei National Laboratory, Hefei, 230088, China

*Corresponding author: yayuwang@tsinghua.edu.cn

In conventional metal superconductors such as aluminum, the large number of weakly bounded Cooper pairs become phase coherent as soon as they start to form. The cuprate high critical temperature ($T_c$) superconductors, in contrast, belong to a distinctively different category. To account for the high $T_c$, the attractive pairing interaction is expected to be strong and the coherence length is short. Being doped Mott insulators[1], the cuprates are known to have low superfluid density, thus are susceptible to phase fluctuations[2]. It has been proposed that pairing and phase coherence may occur separately in cuprates, and $T_c$ corresponds to the phase coherence temperature controlled by the superfluid density[3-5]. To elucidate the microscopic processes of pairing and phase ordering in cuprates, here we use scanning tunneling microscopy to image the evolution of electronic states in underdoped $Bi_2La_xSr_{2-x}CuO_{6+\delta}$. Even in the insulating sample, we observe a smooth crossover from the Mott insulator to superconductor-type spectra on small islands with chequerboard order and emerging quasiparticle interference patterns following the octet model. Each chequerboard plaquette contains approximately two holes, and exhibits a stripy internal structure that has strong influence on the superconducting features. Across the insulator to superconductor boundary, the local spectra remain qualitatively the same while the quasiparticle interferences become long-ranged. These results suggest that the chequerboard plaquette with internal stripes plays a crucial role on local pairing in cuprates, and the global phase coherence is established once its spatial occupation exceeds a threshold.

## Main

The phase fluctuation theory provides a unified picture for explaining the overall phase diagram of cuprates, including the dome-shaped $T_c$ phase line and the pseudogap phase in the underdoped regime[2]. The existence of finite pairing amplitude outside the superconducting dome has been demonstrated by a variety of techniques such as terahertz spectroscopy[6], Nernst effect and diamagnetism[7,8], scanning tunneling microscopy (STM)[9-13], and angle-resolved photoemission spectroscopy (ARPES)[14,15]. But at the microscopic level, it is still unclear how the charge transfer gap of the parent compound evolves into a superconducting gap with increasing doping, and how phase coherence between the dilute, tightly bounded pairs is established. As shown by the $Bi_2Sr_2CuO_{6+\delta}$ (Bi-2201) phase diagram in Fig. 1a, the system remains insulating for hole density up to $p = 0.10$[16], which is a rather high doping level. If the insulator to superconductor quantum phase transition at the ground state is triggered by the formation of long-range phase coherence, the Cooper pairs should already exist for $p < 0.10$. However, so far there is no spectroscopic evidence of local pairing in the insulating regime of underdoped cuprates, and little is known about what happens across the insulator to superconductor phase boundary.

To obtain a microscopic perspective on the pairing and phase ordering processes in cuprates, we use spectroscopic imaging STM to study underdoped Bi-2201 covering both the insulating and superconducting phases. Figure 1b displays the in-plane resistivity ($\rho_{ab}$) vs. temperature ($T$) curves of the three samples studied here, whose doping levels are marked in the phase diagram in Fig. 1a. The $p = 0.08$ sample is highly insulating with a diverging resistivity at low $T$, and the $\rho_{ab}$ value is divided by a factor of 50 to fit into the scale. The $p = 0.11$ sample becomes superconducting with $T_c = 10$ K but it exhibits an insulating behavior between $T_c$ and 80 K, consistent with previous reports on underdoped Bi-2201[17,18]. The $p = 0.13$ sample has a $T_c = 22$ K, and remains metallic for the entire normal state (the estimation of hole density is described in supplementary Session I).

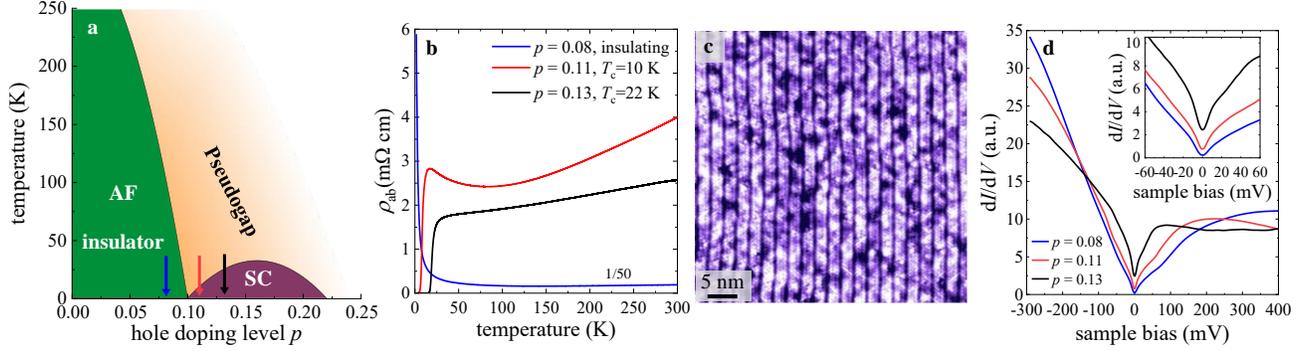

**Fig. 1| Phase diagram and characterization of three underdoped Bi-2201 samples. a**, The schematic phase diagram of Bi-2201. The arrows indicate the hole densities of the three samples studied in this work. **b**, Temperature dependent resistivity of the three samples. The $p = 0.08$ curve is divided by a factor of 50. **c**, Atomically-resolved topographic image acquired at bias voltage $V = -300$ mV and tunneling current $I = 10$ pA on the $p = 0.08$ sample. **d**, Spatially-averaged d$I$/d$V$ spectra of the three samples, where the pseudogap size decreases with increasing doping. Inset: Zoomed-in low energy spectra reveal small gaps with size around 10 meV for all three samples.

Figure 1c shows a topographic image of the $p = 0.08$ sample, in which the structural supermodulation and the surface Bi atoms can be clearly resolved. Figure 1d summarizes the spatially-averaged d$I$/d$V$ spectra of the three samples taken on a large grid, which exhibit systematic evolutions. With increasing doping, the d$I$/d$V$ curves become less asymmetric with respect to the Fermi level ($E_F$), which reflects the weakening of electron correlation effect as the system is doped away from the parent Mott insulator[19]. Meanwhile, as described in supplementary Session II, the average pseudogap size decreases from $\Delta_{PG} \approx 339$ meV at $p = 0.08$ to $\Delta_{PG} \approx 186$ meV and 65 meV for $p = 0.11$ and 0.13, respectively[20]. The two superconducting samples both have a small gap with size $\Delta_{SC} \approx 10$ meV, as shown by the low energy spectra in the inset. Interestingly, the insulating $p = 0.08$ sample also appears to have a shallower gap feature at the same energy scale.

**Similar spectroscopic gaps in the insulating and superconducting samples**

Next, we focus on the spatial evolution of electronic structure in each sample. Figure 2a shows the differential conductance map $g(\mathbf{r}, E) = dI/dV(\mathbf{r}, E)$ taken at $V = 30$ mV on the $p = 0.08$ sample, which visualizes the spatial distribution of electron density of state (DOS) at $E = 30$ meV. There are two pronounced features that are consistent with previous reports on Bi-2201 with slightly lower doping[21]. The first is phase separation into dark areas with nearly undoped Mott insulator spectrum and bright areas with V-shaped pseudogap around $E_F$ (more data are shown in supplementary Session III), which is theoretically expected when dilute holes are dispersed in the antiferromagnetic background[22]. Secondly, the doped charges in the bright areas self-assemble into short-range chequerboard order with wavelength around $4a_0$ ($a_0$ is the lattice constant of the $CuO_2$ plane). The totally new features revealed by this work are displayed in Fig. 2b, when we zoom in the low energy spectra taken along the red line in Fig. 2a. Starting from the dark area (blue curve), there is a complete suppression of low energy DOS and the $dI/dV$ curve is rather featureless. Approaching the edge of the bright area (red curve), the DOS around +300 mV grows significantly due to spectral weight transfer from the high energy Hubbard bands. More interestingly, a shallow, nearly particle-hole symmetric gap around $E_F$ emerges gradually. On top of the bright chequerboard plaquette (magenta curve), the small gap becomes well-developed with coherence peaks around ± 10 mV, highly reminiscent of a superconducting gap.

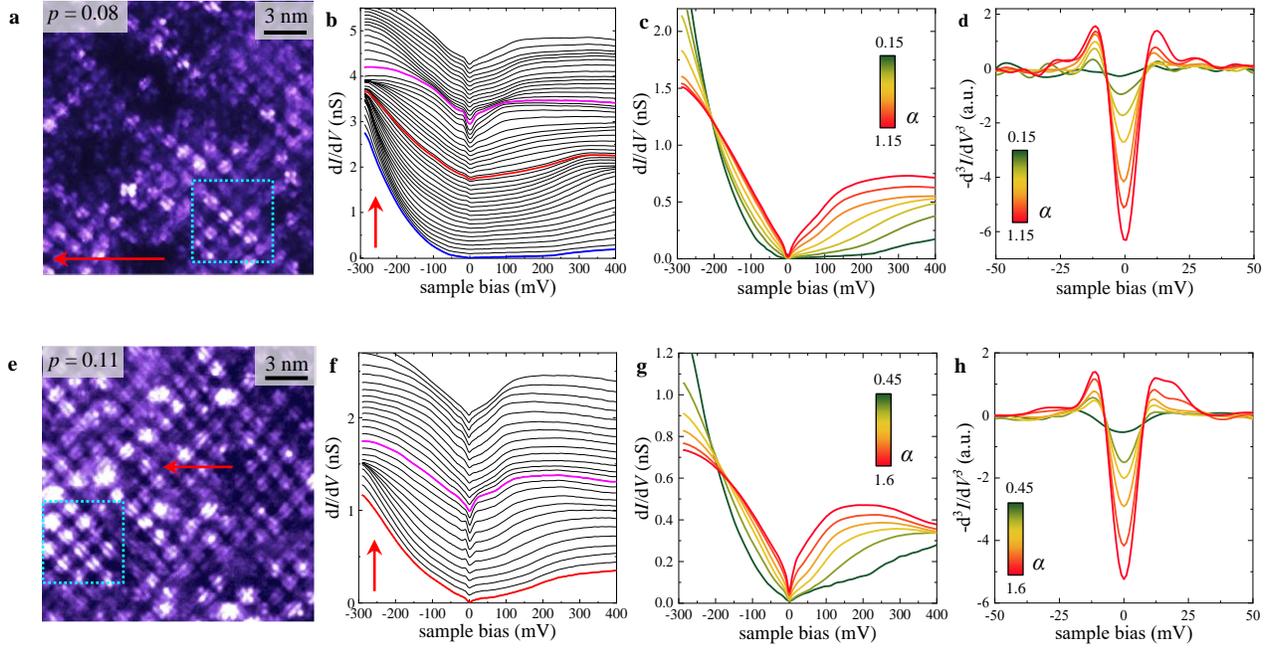

**Fig. 2| Evolution of d$I$/d$V$ spectra on the $p$ = 0.08 and 0.11 samples. a**, The conductance map $g(\mathbf{r}, E = 30\ \text{meV})$ of the $p$ = 0.08 sample. There are dark pits representing the Mott insulator phase, and bright islands of chequerboard order with wavelength around $4a_0$. **b**, The d$I$/d$V$ spectra taken along the red arrow in (a), revealing a continuous evolution from Mott insulator type spectrum to that with a small gap reminiscent of superconducting spectrum. **c**, Clustering analysis of all spectra taken on a dense grid in (a) classified by the Mottness parameter $\alpha$. **d**, The -d$^3I$/d$V^3$ curves of the spectra in (c), showing the emergence of two coherence peaks around $\pm 10$ meV with increasing $\alpha$. **e-h**, The same sets of data taken on the superconducting $p$ = 0.11 sample. The overall behavior is highly similar to that obtained in the bright chequerboard areas in the insulating $p$ = 0.08 sample.

To summarize the spectral evolution over the whole sample, in Fig. 2c we use clustering analysis to classify the 241×241 spectra taken on a dense grid in Fig. 2a into 7 curves according to the Mottness parameter defined as $\alpha = \int_{-0.1\ \text{eV}}^{0.4\ \text{eV}} g(E)\text{d}E / \int_{-0.3\ \text{eV}}^{-0.1\ \text{eV}} g(E)\text{d}E$, which characterizes the spectral weight transferred into the charge transfer gap[23] (definition of $\alpha$ is explained in supplementary Session IV). With increasing $\alpha$, hence local hole density, there is a continuous decrease of tunneling asymmetry as well as the pseudogap size. Along with this process, the small gap around $E_F$ becomes deeper and sharper. To enhance the low energy features, we plot the second derivative of d$I$/d$V$, or -d$^3I$/d$V^3$, in Fig. 2d. For the curve with the smallest $\alpha$, -d$^3I$/d$V^3$ remains zero within the noise floor, consistent with the expectation for undoped Mott insulator. With increasing $\alpha$, a zero bias dip and two peaks around ± 10 mV emerge gradually and become sharp and symmetric for $\alpha > 1$.

Figure 2e is the d$I$/d$V$ map taken at $V = 30$ mV on the superconducting $p = 0.11$ sample. The chequerboard pattern becomes more uniform, and the dark pits of Mott insulating phase are small and sparse. Figure 2f displays the d$I$/d$V$ curves taken along the red line in Fig. 2e, which exhibit systematic evolutions resembling the upper half of Fig. 2b above the red curve. Approaching the bright chequerboard area, the pseudogap size reduces continuously and the small gap features become more pronounced. The clustering analysis classification of the spectra (Fig. 2g) and the corresponding second derivative plots (Fig. 2h) are also qualitatively similar to that in the $p = 0.08$ sample, except that here the small energy gap is present even for the smallest $\alpha$. Because this sample is already in the superconducting state at the measurement temperature $T = 4.7$ K, there is no doubt that the peaks in Fig. 2h are the coherence peaks of a superconducting gap with $\Delta_{SC} \approx 9$ meV. Therefore, it is reasonable to assume that the nearly identical low energy spectra on the bright chequerboard areas of the $p = 0.08$ sample are also due to a superconducting gap when the local hole density is relatively high.

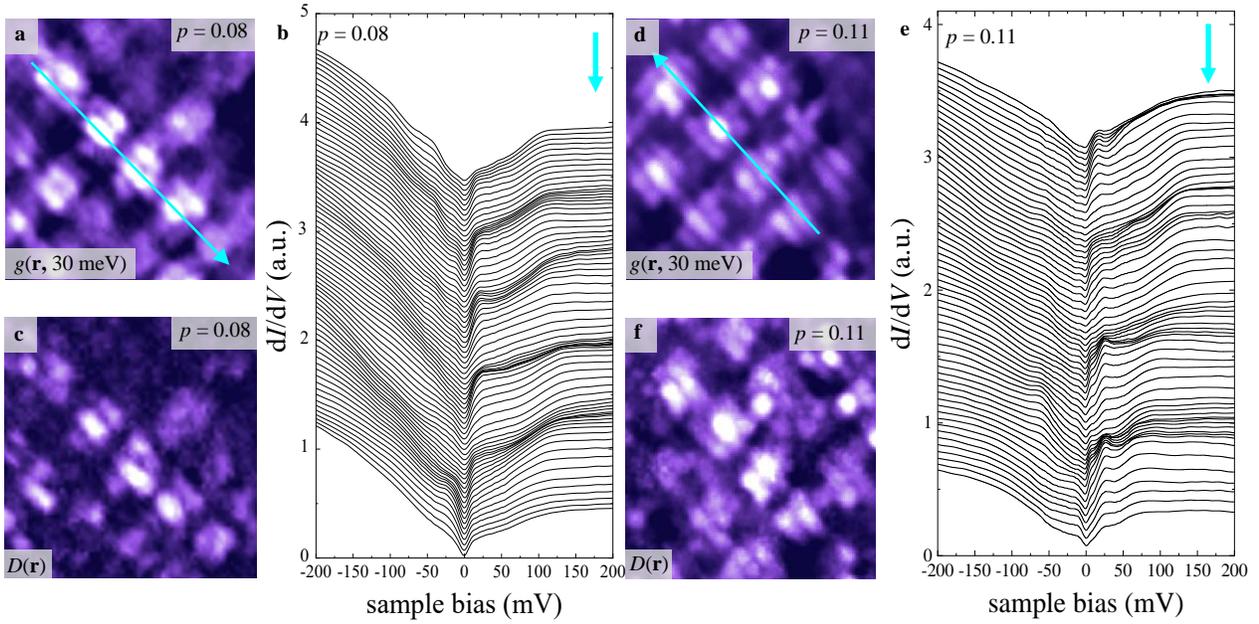

**Fig. 3| Periodic gap features on the $p$ = 0.08 and 0.11 samples. a** and **d**, Zoomed-in d$I$/d$V$ maps on the areas enclosed by the dashed cyan boxes in Fig. 2a and 2e. **b** and **e**, d$I$/d$V$ curves along the cyan arrows in **a** and **e** on the $p$ = 0.08 and 0.11 samples, where the spectra exhibit spatial variations with the same periodicity as the chequerboard patterns. **c** and **f**, The superconducting coherence peak sharpness map $D(\mathbf{r})$, revealing similar periodic chequerboard and internal stripe patterns.

The similarity between the insulating and superconducting samples is reinforced by the same periodic modulations of the small gap features. Shown in Fig. 3b are the d$I$/d$V$ spectra taken along the cyan line in Fig. 3a over a bright area of the $p$ = 0.08 sample with regular chequerboard pattern. The spectral lineshape modulates periodically in space and reveal more pronounced coherence peaks for curves taken on top of the bright chequerboard plaquette. Figure 3c displays the peak sharpness map $D(\mathbf{r})^{24,25}$, extracted from the second derivatives -d$^3I$/d$V^3$($\mathbf{r}$, $V_{sc}$) at the coherence peak energy. It reveals the chequerboard periodicity and stripy internal structure of the coherence peak. Figures 3d-f are the same sets of data for the $p$ = 0.11 sample, which show very similar spatial variations. In fact, if we only compare these two sets of data, it is difficult to distinguish the insulating sample from the superconducting one. Such behavior has been observed before in an underdoped $Bi_2Sr_2CaCu_2O_{8+\delta}$ (Bi-2212) with $T_c = 10$ K, in which the coherence peaks and depth of the superconducting gap modulate periodically with the same wavelength as the chequerboard[24]. The d$I$/d$V$ spectra on the bright stripes have more pronounced coherence peaks, as revealed by the $D(\mathbf{r})$ map in Fig. 3c, which has also been observed in overdoped superconducting Bi-2201[25]. These results provide further supports that local pairs already exist in the bright chequerboard area of the $p = 0.08$ insulator. ARPES and thermal transport studies of other cuprates revealed similar nodal excitations straddling the superconductor-insulator transition point[26,27], and the diamagnetization experiment observed the persistence of vortex liquid phase to lightly doped insulating regime[8]. These results support the picture that Cooper pairs already exist before the system becomes superconducting. Moreover, the coexistence of charge glass and superconducting fluctuation in the insulating side implied from macroscopic transport measurement[28] is also consistent with the concurrence of superconducting gap and chequerboard plaquette directly visualized here.

**Long-range phase coherence across the insulator-superconductor transition**

To visualize what happens across the insulator to superconductor boundary, we then zoom out to a larger scale to gain a global view. Figures 4a-c are the tunneling current map $I(\mathbf{r}, 30$ meV$)$ over a larger field of view ($37 \times 37$ nm$^2$) for the three samples. Because $I(\mathbf{r}, 30$ meV$)$ is proportional to the integrated local DOS from $E_F$ to 30 meV, it averages out the bias dependent dispersive features and enhances the non-dispersive static order. All three maps exhibit clear chequerboard patterns, but there is a crucial difference: the spatial occupation of chequerboard plaquette varies significantly. For the $p = 0.08$ sample, the chequerboard regions are separated into small islands surrounded by dark Mott insulating phase. For the $p$ = 0.11 sample, the chequerboard plaquettes are more uniformly distributed and cover the vast majority of the surface. The chequerboard plaquettes in the $p$ = 0.13 sample cover the whole surface, giving an overall brighter background and a significant number of very bright spots. To give a quantitative estimate of the chequerboard occupation, we count the number of plaquettes in Fig. 4a-c (see supplementary Session V for details), which has a density of $0.040/a_0^2$, $0.053/a_0^2$, and $0.045/a_0^2$, respectively. Remarkably, by dividing the hole density by the plaquette density, we find that on average each plaquette contains 1.99 and 2.09 holes for the $p$ = 0.08 and 0.11 samples. The Fourier transform images of Fig. 4a-c reveal the chequerboard wavevectors along the Cu-O bond direction, and the wavelength can be extracted to be $3.9a_0$, $4.0a_0$, and $4.9a_0$, respectively (see supplementary Session VI for details). This trend is consistent with previous report that the charge order wavelength remains $\approx 4a_0$ for $p$ < 0.12, and increases at higher doping[29-32].

Figures 4d-f display the conductance ratio map $Z(\mathbf{r}, 10\ \text{meV}) = g(\mathbf{r}, 10\ \text{meV})/g(\mathbf{r}, -10\ \text{meV})$, which is an effective way to enhance the dispersive quasiparticle interference (QPI) features[10,33], on the same field of view. For the $p$ = 0.08 sample, the main features are large bright and dark patches, but there are also weak wave-like dispersions in the region with sizable chequerboard order. Its Fourier transform image in Fig. 4j reveals rather fuzzy but certainly discernable Bogoliubov QPI patterns that roughly follow the octet model[34,35]. It is clear evidence for the existence of Cooper pairing with short-range phase correlations on chequerboard islands in insulating cuprate (see supplementary Session VII for more QPI results on this sample). Figure 4h-i display the QPI patterns in the $p$ = 0.11 and 0.13 samples, respectively, in which the wavevectors become more well-defined and perfectly match the octet model for $d$-wave superconductivity,

implying the establishment of long-range phase coherence. Figures 4m and 4n summarize the QPI wavevectors at different energies, indicating that the QPI gets more dispersive and closer to $d$-wave symmetry with increasing doping level. It is consistent with the observation of persistent octet-model QPI above $T_c$ in underdoped cuprates, and the picture that phase fluctuations make the Bogoliubov QPI less dispersive[9,36,37]. Here by reducing hole density in the ground state, the long-range phase coherence is lost across the superconductor to insulator phase boundary, but local Cooper pairing with short-range phase correlation and less dispersive Bogoliubov QPs persist in the sizable chequerboard islands.

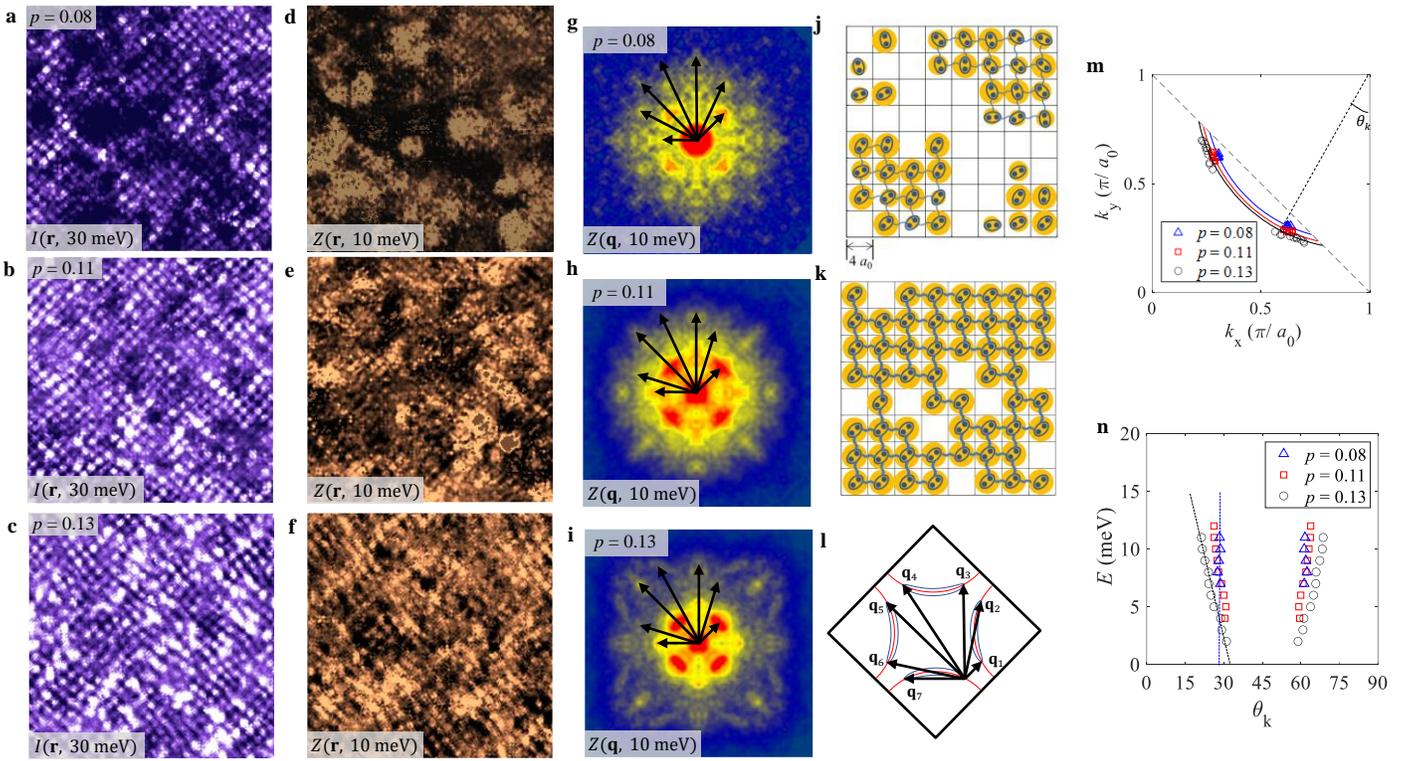

**Fig. 4| Quasiparticle interferences and establishment of long-range phase coherence across the insulator-superconductor phase boundary. a-c**, The tunneling current maps $I(\mathbf{r}, 30 \text{ meV})$ on the three samples. **d-f**, Conductance ratio maps $Z(\mathbf{r}, 10 \text{ meV})$ on the three samples, where dispersive QPI features are present in the two superconducting samples and the chequerboard area of insulating sample. **g-i**, The

Fourier transform images reveal the emergence and growth of QPI with increasing doping. **j**, Schematic diagram of local pairs with short-range phase correlations on disconnected chequerboard islands without long-range phase coherence in the insulating phase. **k**, Schematic diagram showing the formation of long-range phase coherence when the chequerboard density exceeds a threshold. **l**, Schematic $k$-space structure of the octet model for QPI in a $d$-wave superconductor with the seven wavevectors matching that in **i**. **m**, The QPI wavevector extracted from the octet pattern at different energies. The QPs are more dispersive in the higher doping level sample. The arcs are the Fermi surfaces calculated by the tight-binding model for $p$ = 0.08, 0.11 and 0.13, respectively. **n**. The energy dependence of the angel of QPI wavevectors. The energy dependence of scattering wavevectors gets closer to $d$-wave symmetry with increasing doping level.

Based on the observations described above, we arrive at a consistent physical picture regarding the emergence of local pairing and global phase coherence in underdoped cuprates. When a few percent of holes are doped into the parent Mott insulator, they spontaneously segregate into islands of chequerboard order with plaquette size around $4a_0$. Both the small gap and QPI features on the bright chequerboard areas are strikingly similar to that of the superconducting phase, indicating the formation of local pairing with short-range phase correlations. Due to the lack of long-range phase coherence between the chequerboard islands separated by Mott insulating regions, the system behaves like an insulator, as shown schematically in Fig. 4j. With increasing doping, the chequerboard islands grow continuously and cover a larger portion of the sample. When its spatial occupation exceeds a threshold, long-range phase coherence between the chequerboard islands is established and the system becomes a superconductor, as shown schematically in Fig. 4k. The more uniform superfluid density leads to highly dispersive Bogoliubov QPI patterns with well-defined wavevectors that can be described by the octet model illustrated in Fig. 4l.

The insulator to superconductor quantum phase transition, as dramatic as it appears, is caused by the coalescing of chequerboard islands with the quantitative increase of plaquette occupation. The global phase coherence is established in two steps, first between the local pairs within a chequerboard island, and then long-range phase coherence between islands. Phenomenologically, it is consistent with the theoretical model of Josephson coupling induced phase locking between superconducting clusters[38,39], which has been proposed to explain the relation between phase stiffness and $T_c$ of underdoped cuprate[3,4,40]. An important new insight revealed by our results is that at the microscopic level, the elementary building blocks that eventually form the phase coherent condensate are islands of regular chequerboard plaquettes, rather than randomly distributed superconducting clusters due to disorders[41-43]. A more fundamental question is why the doped holes tend to form chequerboard plaquettes with localized pairs in the first place. Naively, it must be related to the minimization of electronic energy, like that described by the stripe model for doped antiferromagnetic Mott insulator[44-47]. Recently, pairing induced by strong electron correlation in lightly doped *t-J* model has been shown numerically[48,49]. Here we directly visualized that the lowest energy state is achieved by the spontaneous formation of chequerboard with approximately two holes on a plaquette with size around $4a_0$ and internal stripes, where local pairing first emerges. These microscopic structures may hold the key for unveiling the pairing mechanism of high $T_c$ superconductivity in cuprates.

## Methods

**Sample growth.** The lanthanum substituted Bi-2201 samples studied in this work are grown by the floating zone method[15] with nominal La ingredient 0.84 in the $p = 0.08$ and 0.11 samples, and La ingredient 0.6 in the $p = 0.13$ sample. Different post annealing treatments are used to improve the sample quality and vary the doping levels.

**STM/STS measurements.** The STM experiments are carried out on an ultrahigh vacuum low-temperature system. The samples are cleaved *in situ* at 77 K with base pressure lower than $1.0 \times 10^{-10}$ mbar, and immediately transferred into the STM head cooled at $T = 4.7$ K. The topographic images are collected in the constant current mode and the differential conductance spectra are acquired by a standard lock-in method with $f = 723.137$ Hz.

## Data availability

All data are available in the main text or the supplementary materials.

## References

1   Lee, P. A., Nagaosa, N. & Wen, X.-G. Doping a Mott insulator: Physics of high-temperature superconductivity. *Rev. Mod. Phys.* **78**, 17-85 (2006).
2   Emery, V. & Kivelson, S. Importance of phase fluctuations in superconductors with small superfluid density. *Nature* **374**, 434-437 (1995).
3   Uemura, Y. *et al.* Universal correlations between $T_c$ and $n_s/m^*$ (carrier density over effective mass) in high-$T_c$ cuprate superconductors. *Phys. Rev. Lett.* **62**, 2317 (1989).
4   Homes, C. *et al.* A universal scaling relation in high-temperature superconductors. *Nature* **430**, 539-541 (2004).

5    Bozovic, I., He, X., Wu, J. & Bollinger, A. T. Dependence of the critical temperature in overdoped copper oxides on superfluid density. *Nature* **536**, 309-311 (2016).

6    Corson, J., Mallozzi, R., Orenstein, J., Eckstein, J. & Bozovic, I. Vanishing of phase coherence in underdoped $Bi_2Sr_2CaCu_2O_{8+\delta}$. *Nature* **398**, 221-223 (1999).

7    Wang, Y. *et al.* Field-Enhanced Diamagnetism in the Pseudogap State of the Cuprate $Bi_2Sr_2CaCu_2O_{8+\delta}$ Superconductor in an Intense Magnetic Field. *Phys. Rev. Lett.* **95**, 247002 (2005).

8    Li, L., Checkelsky, J., Komiya, S., Ando, Y. & Ong, N. Low-temperature vortex liquid in $La_{2-x}Sr_xCuO_4$. *Nat. Phys.* **3**, 311-314 (2007).

9    Lee, J. *et al.* Spectroscopic fingerprint of phase-incoherent superconductivity in the underdoped $Bi_2Sr_2CaCu_2O_{8+\delta}$. *Science* **325**, 1099-1103 (2009).

10   Kohsaka, Y. *et al.* How Cooper pairs vanish approaching the Mott insulator in $Bi_2Sr_2CaCu_2O_{8+\delta}$. *Nature* **454**, 1072-1078 (2008).

11   Gomes, K. K. *et al.* Visualizing pair formation on the atomic scale in the high-$T_c$ superconductor $Bi_2Sr_2CaCu_2O_{8+\delta}$. *Nature* **447**, 569-572 (2007).

12   Pasupathy, A. N. *et al.* Electronic origin of the inhomogeneous pairing interaction in the high-$T_c$ superconductor $Bi_2Sr_2CaCu_2O_{8+\delta}$. *Science* **320**, 196-201 (2008).

13   Kugler, M., Fischer, O., Renner, C., Ono, S. & Ando, Y. Scanning tunneling spectroscopy of $Bi_2Sr_2CuO_{6+\delta}$: new evidence for the common origin of the pseudogap and superconductivity. *Phys. Rev. Lett.* **86**, 4911-4914 (2001).

14   Loeser, A. *et al.* Excitation gap in the normal state of underdoped $Bi_2Sr_2CaCu_2O_{8+\delta}$. *Science* **273**, 325-329 (1996).

15   Ding, H. *et al.* Spectroscopic evidence for a pseudogap in the normal state of underdoped high-$T_c$ superconductors. *Nature* **382**, 51-54 (1996).

16   Peng, Y. *et al.* Disappearance of nodal gap across the insulator-superconductor transition in a copper-oxide superconductor. *Nat. Commun.* **4**, 2459 (2013).

17   Ando, Y. *et al.* Metallic in-plane and divergent out-of-plane resistivity of a high-$T_c$ cuprate in the zero-temperature limit. *Phys. Rev. Lett.* **77**, 2065 (1996).

18   Ando, Y. *et al.* Normal-state Hall effect and the insulating resistivity of high-$T_c$ cuprates at low temperatures. *Phys. Rev. B* **56**, R8530 (1997).

19  Anderson, P. W. & Ong, N. P. Theory of asymmetric tunneling in the cuprate superconductors. *J. Phys. Chem. Solids* **67**, 1-5 (2006).

20  Kohsaka, Y. *et al.* Visualization of the emergence of the pseudogap state and the evolution to superconductivity in a lightly hole-doped Mott insulator. *Nat. Phys.* **8**, 534-538 (2012).

21  Cai, P. *et al.* Visualizing the evolution from the Mott insulator to a charge-ordered insulator in lightly doped cuprates. *Nat. Phys.* **12**, 1047-1051 (2016).

22  Emery, V. J., Kivelson, S. A. & Lin, H. Q. Phase separation in the *t-J* model. *Phys. Rev. Lett.* **64**, 475-478 (1990).

23  Battisti, I. *et al.* Universality of pseudogap and emergent order in lightly doped Mott insulators. *Nat. Phys.* **13**, 21-25 (2016).

24  Ruan, W. *et al.* Visualization of the periodic modulation of Cooper pairing in a cuprate superconductor. *Nat. Phys.* **14**, 1178-1182 (2018).

25  Li, X. *et al.* Evolution of Charge and Pair Density Modulations in Overdoped $Bi_2Sr_2CuO_{6+\delta}$. *Phys. Rev. X* **11**, 011007 (2021).

26  Chatterjee, U. *et al.* Observation of a d-wave nodal liquid in highly underdoped $Bi_2Sr_2CaCu_2O_{8+\delta}$. *Nat. Phys.* **6**, 99-103 (2009).

27  Doiron-Leyraud, N. *et al.* Onset of a Boson mode at the superconducting critical point of underdoped $YBa_2Cu_3O_y$. *Phys. Rev. Lett.* **97**, 207001 (2006).

28  Shi, X. *et al.* Emergence of superconductivity from the dynamically heterogeneous insulating state in $La_{2-x}Sr_xCuO_4$. *Nat. Mat.* **12**, 47-51 (2013).

29  Peng, Y. Y. *et al.* Direct observation of charge order in underdoped and optimally doped $Bi_2(Sr,La)_2CuO_{6+\delta}$ by resonant inelastic X-ray scattering. *Phys. Rev. B* **94**, 184511 (2016).

30  Comin, R. *et al.* Charge order driven by Fermi-arc instability in $Bi_2Sr_{2-x}La_xCuO_{6+\delta}$. *Science* **343**, 390-392 (2014).

31  Wise, W. *et al.* Charge-density-wave origin of cuprate checkerboard visualized by scanning tunnelling microscopy. *Nat. Phys.* **4**, 696-699 (2008).

32  Webb, T. A. *et al.* Density wave probes cuprate quantum phase transition. *Phys. Rev. X* **9**, 021021 (2019).

33  Hanaguri, T. *et al.* Quasiparticle interference and superconducting gap in $Ca_{2-x}Na_xCuO_2Cl_2$. *Nat. Phys.* **3**, 865-871 (2007).

34  Wang, Q.-H. & Lee, D.-H. Quasiparticle scattering interference in high-temperature superconductors. *Phys. Rev. B* **67**, 020511 (2003).

35  Hoffman, J. *et al.* Imaging quasiparticle interference in $Bi_2Sr_2CaCu_2O_{8+\delta}$. *Science* **297**, 1148-1151 (2002).

36  Chubukov, A. V., Norman, M. R., Millis, A. J. & Abrahams, E. Gapless pairing and the Fermi arc in the cuprates. *Phys. Rev. B* **76**, 180501 (2007).

37  Wulin, D., He, Y., Chien, C.-C., Morr, D. K. & Levin, K. Model for the temperature dependence of the quasiparticle interference pattern in the measured scanning tunneling spectra of underdoped cuprate superconductors. *Phys. Rev. B* **80**, 134504 (2009).

38  Imry, Y., Strongin, M. & Homes, C. An inhomogeneous Josephson phase in thin-film and high-$T_c$ superconductors. *Physica C: Superconductivity* **468**, 288-293 (2008).

39  Imry, Y., Strongin, M. & Homes, C. $n_s$−$T_c$ Correlations in Granular Superconductors. *Phys. Revi. Lett.* **109**, 067003 (2012).

40  Zuev, Y., Kim, M. S. & Lemberger, T. R. Correlation between superfluid density and $T_c$ of Underdoped $YBa_2Cu_3O_{6+x}$ near the superconductor-insulator transition. *Phys. Rev. Lett.* **95**, 137002 (2005).

41  Haviland, D., Liu, Y. & Goldman, A. Onset of superconductivity in the two-dimensional limit. *Phys. Rev. Lett.* **62**, 2180 (1989).

42  Yazdani, A. & Kapitulnik, A. Superconducting-insulating transition in two-dimensional *α*-MoGe thin films. *Phys. Rev. Lett.* **74**, 3037 (1995).

43  Mason, N. & Kapitulnik, A. Dissipation effects on the superconductor-insulator transition in 2D superconductors. *Phys. Rev. Lett.* **82**, 5341 (1999).

44  Zaanen, J. & Gunnarsson, O. Charged magnetic domain lines and the magnetism of high-$T_c$ oxides. *Phys. Rev. B* **40**, 7391 (1989).

45  Tranquada, J., Sternlieb, B., Axe, J., Nakamura, Y. & Uchida, S. Evidence for stripe correlations of spins and holes in copper oxide superconductors. *Nature* **375**, 561-563 (1995).

46  Kivelson, S. A., Fradkin, E. & Emery, V. J. Electronic liquid-crystal phases of a doped Mott insulator. *Nature* **393**, 550-553 (1998).

47  Tranquada, J. M., Dean, M. P. & Li, Q. Superconductivity from charge order in cuprates. *J. Phys. Soc. Jpn.* **90**, 111002 (2021).

48  Jiang, H.-C. & Kivelson, S. A. Stripe order enhanced superconductivity in the Hubbard model. *Proceedings of the National Academy of Sciences* **119**, e2109406119 (2022).

49  Zhao, J.-Y., Chen, S. A., Zhang, H.-K. & Weng, Z.-Y. Two-Hole Ground State: Dichotomy in Pairing Symmetry. *Phys. Rev. X* **12**, 011062 (2022).

## Acknowledgments

We thank Z.Y. Weng, G.M. Zhang, and C.L. Song for helpful discussions and technical support. This work was supported by the Basic Science Center Project of NSFC under grant No. 52388201, NSFC Grant No. 11888101. Y.W. is supported by the Innovation Program for Quantum Science and Technology (grant No. 2021ZD0302502), and the New Cornerstone Science Foundation through the New Cornerstone Investigator Program and the XPLORER PRIZE. X.J.Z. is supported by the Strategic Priority Research Program (B) of the Chinese Academy of Sciences (XDB25000000).

## Author contributions

Y.W. and X.J.Z. supervised this project. H.Y. and Y.C. prepared the single crystal. S.Y., C.Z. and M.X. carried out the STM experiments under the supervision of Y.W. Z.H., Y.J., and S.Y. performed the transport experiments. S.Y. and Y.W. prepared the manuscript with comments from all authors.

## Competing interests

The authors declare no competing interests.